\documentclass{article}

\usepackage{arxiv}

\usepackage[utf8]{inputenc} 
\usepackage[T1]{fontenc}    
\usepackage{hyperref}       
\usepackage{url}            
\usepackage{booktabs}       
\usepackage{amsfonts}       
\usepackage{nicefrac}       
\usepackage{microtype}      
\usepackage{amsmath}
\usepackage{float}
\usepackage{varioref}
\usepackage[makeroom]{cancel}
\usepackage{graphicx}

\newcommand{\Mathematica}{\textit{Mathematica\textsuperscript{\resizebox{!}{0.8ex}{\textregistered}}}}

\title{Analytic and Numerical Study Of Navier-Stokes Loop Equation in Turbulence}

\author{
  Alexander Migdal \\
  Fresnel Research LLC
  } 
\begin{document}
\maketitle

\begin{abstract}
We developed analytic approach to the non-planar loop equation, which we derived in previous papers \cite{M19a},\cite{M19b},\cite{M19c}. We found quadratic integral equation for the vorticity distribution $\Omega(r)$ we introduced on a minimal surface.  There are no corrections to the minimal surface though: it is still defined by mean external curvature equal to zero, for arbitrary non-planar loop. We also analyzed the loop equations with viscosity term in Navier-Stokes equations. This term creates boundary condition for $\Omega(r\in C)$. The leading viscosity correction term mixes the moments $\left< \Gamma^p \right>$  with $\left< \Gamma^{p-1} \right>$ resembling the bi-fractal behavior observed in \cite{S19} and explicitly breaking the time reversal symmetry. We also develop numerical approach to the loop equation with arbitrary curved loop and present \Mathematica notebook building triangulated minimal surface and then numerically solving these equations. As a result we obtain predictions for future numerical experiments which will compute vorticity distribution along the loop.

\end{abstract}


\section{Introduction}
In the previous papers \cite{M19a,M19b,M19c} we revived and advanced the old conjecture \cite{M93} that tails of velocity circulation PDF in inertial range of developed turbulence are determined by a minimal surface bounded by a given loop. This conjecture was (to some extent) confirmed in recent numerical experiments  \cite{S19}, which inspired this renewed interest to the Loop Equations of \cite{M93}.

The basic variable in the Loop Equations a circulation around closed loop in coordinate space

\begin{equation}\label{Gamma}
    \Gamma[C] = \oint_C \vec v d\vec r 
\end{equation}

The PDF  for  velocity circulation as a functional of the loop 
\begin{equation}
    P\left ( C,\Gamma\right) =\left < \delta\left(\Gamma - \oint_C \vec v d\vec r\right)\right>
\end{equation}

with brackets \begin{math}< > \end{math} corresponding to time average or average over random forces, was shown to satisfy certain functional equation (loop equation). 

We shall reproduce the basic part of previous results on a new level of understanding we achieved since writing those papers.

Let us summarize the physical picture before we get into mathematical details of the loop equations.
We are looking for some spatial distribution of vorticity which would preserve the circulation in the Navier-Stokes equations
\begin{equation}
    \dot{v}_i = \nu \partial_j \partial_j v_i + v_j \partial_j v_i - \partial_i p;\, \partial_i v_i =0
\end{equation}
Differentiating the circulation and dropping total derivative terms in the closed loop integral we get Kelvin-Helmholtz equaton
\begin{align}\label{KH}
     \oint_C d r_i  \dot{v}_i &= \nu \oint_C d r_i  \partial_j \left(\partial_j v_i - \partial_i v_j\right) + \oint_C d r_i v_j \left(\partial_j v_i - \partial_i v_j\right) \\
     &=  \nu e_{j i k} \oint_C d r_i  \partial_j \omega_k + e_{j i k}\oint_C d r_i v_j \omega_k 
\end{align}

The velocity here is related to vorticity by Biot-Savart law
\begin{equation}
    v_a(r) = e_{a b c} \frac{\partial}{\partial r_b} \int d^3 r' \frac{\omega_c(r')}{4 \pi |\vec r'- \vec r|} 
\end{equation}

The loop equation follows from the Kelvin-Helmholtz equation and Biot-Savart law:
\begin{align}\label{LoopEq0}
&\partial_\Gamma \partial_t  P\left ( C,\Gamma\right) 
= \\
&\oint_C d r_i \nu e_{i j k} \frac{\partial}{\partial r_j} \partial_\Gamma \frac{\delta P(C,\Gamma)}{\delta \sigma_k(r)} +\\
&\oint_C d r_i \int d^3 r'\frac{\delta^2 P(C,\Gamma)}{\delta \sigma_k(r) \delta \sigma_l(r')}\left(\delta_{k l}\frac{\partial}{\partial r_i} - \delta_{i l}\frac{\partial}{\partial r_k}\right)\frac{1 }{4\pi|\vec r - \vec r'|}
\end{align}
The definition and the meaning of the area derivative $\frac{\delta }{\delta \sigma_k(r)}$ will be explained in the next section.

We are going to find the specific distribution of vorticity field which leads to conservation of circulation \textbf{for this particular loop}. This corresponds to stationary solution of the loop equation, where the right side vanishes.

Let us stress an important distinction between generic stationary solution of the Navier-Stokes equations and what we are looking for here.  We are investigating the PDF of velocity circulation as a function of circulation and a functional of the shape of the loop. So, we are selecting among all possible histories of Navier-Stokes evolution in presence of random forces only those with fixed circulation $\Gamma$ around specific contour $C$, or, in terms of probability, we are studying conditional probability density for velocity circulation.

So, the dominant velocity/vorticity spatial distribution providing this stationary circulation is not the same as unrestricted stationary solution of NS equations, not just because of random forces but also because of extra condition on probability distribution. We assume that the circulation $\Gamma$ is much larger than viscosity, and the "classical" part of vorticity is much larger than its fluctuation. We are looking for this "mean field" which will determine the PDF tails in the same way as instantons do in quantum field theory and stochastic PDE \cite{FKLM}.

One more important point. Initially, in the old papers \cite{M93} we only claimed area law as an asymptotic law for the tails of PDF. Recently, we observed that for 2D fluid Euler dynamics it satisfied the loop equations exactly, not just for large circulations. 

Here, in the context of nonplanar loop equation (which is generic case in more than two dimensions) we shall see that the minimal surface solution holds only as an asymptotic solution for the tails of PDF. This is the WKB approximation to the loop equation, corresponding to the instanton on loop space we mentioned above.

So, we are looking for the loop space instanton in Navier-Stokes equation to determine the circulation PDF tails.
Why loop space? Because the loop equations appear to be the unique non-perturbative method to study fluctuating vector fields in statistical field theory. They were extensively studied in the context of QCD where they are in fact more complex because of non-Abelian gauge field. Some closed form solutions were found there, but only in form of infinite matrix models or infinite algebras. The fluid dynamics represents the Abelian velocity field which simplifies the loop equation. On the other hand, the dynamics is nonlocal because vorticity generates velocity filed all over space by Biot-Savart law. This is the main distinction of fluid dynamics from Abelian gauge theory such as QED.

Let come back to Navier-Stokes instanton and us try thin viscosity sheet bounded by the loop $C$. 
In general the loop is not flat, and the sheet could potentially be any shape. Let us further demand that vorticity is directed towards the normal $n_i(r)$ to the sheet at every point.  
\begin{equation}
    \omega_i(r) \propto n_i(r)
\end{equation}
How can this be compatible with conservation law
\begin{equation}
    \partial_i \omega_i =0
\end{equation}
which follows from the fact that vorticity is a curl of velocity? Let us study this issue in some detail.

In this paper we shall use the notation for the derivatives of a function along the surface $\vec r = \vec S(u,v)$:
\begin{equation}
    \partial_i = g^{\alpha \beta} \partial_{\beta} S^i\partial_{\alpha}
\end{equation}
where Latin indexes $i =1,2,3$ refer to 3D space and Greek indexes $\alpha = 1,2$ refer to internal coordinates$u,v$ on the surface. The induced metric tensor
\begin{equation}
    g_{\alpha \beta} = \partial_{\alpha} S^i \partial_{\beta} S^i
\end{equation}
and $g^{\alpha \beta} $ is an inverse matrix as usual.
The normal vector
\begin{equation}
    n_i \propto e_{i j k} \partial_1 S^j \partial_2 S^k 
\end{equation}
with normalization factor such that $n_i n_i =1$.

One can check that 
\begin{equation}
    \partial_{\alpha} S^i \partial_i = \partial_{\alpha}; 
\end{equation}
so that in particular the line integration of a gradient is covariant
\begin{equation}
    d S_i \partial_i \phi = d \xi_{\alpha} \partial_{\alpha} \phi
\end{equation}
Also, obviously
\begin{equation}
    n_i \partial_i =0
\end{equation}
because the normal vector is orthogonal to both repers $\partial_{\alpha} S^i$.
Once applied to the function of the point $r$ in space as projected on the surface we have
\begin{align}
    &\partial_i F\left(S(\xi)\right) = \left[P^{i k}(r) \frac{\partial F(r)}{\partial r_k}\right]_{r = S(\xi)};\\
    & P^{i k}(S) = \partial_{\alpha} S^i \partial^{\alpha} S^k = \delta_{i k} - n_i(r) n_k(r)
\end{align}
This projection tensor selects derivatives along the surface, killing the derivatives in normal direction. We shall use these projection tensors when needed to convert spatial derivatives to the directions along the surface.

Let us verify $\partial_i n_i=0$.
In linear vicinity of local tangent plane to the surface its equation reads ( with $K_1, K_2$ being principal curvatures at this point)
\begin{align}
  & z - \frac{K_1}{2} x^2 - \frac{K_2}{2} y^2 =0\\
  &n_i = \frac{(-K_1 x , -K_2 y, 1)}{\sqrt{1 + K_1^2 x^2 + K_2^2 y^2}} \rightarrow (-K_1 x , -K_2 y, 1)\\
  &\partial_{\alpha} S^i \rightarrow \delta_{\alpha i} + \delta_{3 i} K_{\alpha} x_{\alpha} \rightarrow \delta_{\alpha i}\\
  &\partial_i \rightarrow (\partial_x, \partial_y, 0)
\end{align}
therefore the conservation law requires
\begin{equation}
    \partial_i n_i \rightarrow -K_1 - K_2 =0
\end{equation}
which is nothing but an equation for the minimal surface.
We can also find the full matrix of derivatives of the normal vector on a minimal surface:
\begin{equation}\label{PartialOmega}
    \partial_j n_k = K_{j k l}  n_l; 
\end{equation}
where in the coordinate frame where $ n= (0,0,1)$
\begin{equation}
    K_{j k 3} = K_1\left(\delta_{j 2} \delta_{k 2} - \delta_{j 1} \delta_{k 1} \right)
\end{equation}

In general coordinate frame
\begin{equation}
     K_{j k l}= -\partial^\alpha S^j \partial^\beta S^k\partial_\alpha \partial_\beta S^l
\end{equation}
and the mean curvature equation in general frame reads
\begin{equation}
    K_{j j l} n_l = -n_l \partial^\alpha \partial_\alpha S^l =0
\end{equation}
Let us disregard the viscosity term and study Euler Loop Equation in the inertial range.\footnote{We are going to study viscosity term below and we find our that it is not in fact negligible, as it imposes boundary condition on vorticity field in our instanton.}
Assuming vorticity to be spread in thin layer of thickness $r_0$ around minimal surface $S_C$ we get from the Biot-Savart integral
\begin{equation}
    e_{j i k}\oint_C d r_i v_j \omega_k \propto  r_0 \left(\delta_{i b} \delta_{k c} - \delta_{i c} \delta_{k b}\right)\oint_C d r_i \omega_k(r)\int_{S_C} d\sigma(r') \frac{r'_b- r_b}{4 \pi |\vec r'- \vec r|^3} \omega_c(r')
\end{equation}

Now, this integral does not vanish by itself, and we do not have free variable to make it vanish. 

So, we generalize the Anzatz, by introducing local vorticity strength at every point of minimal surface
\begin{equation}
    \omega_i(r) =  n_i(r) \Omega(r)
\end{equation}
If this new field varies only along the surface but not in the normal direction, vorticity will still be conserved
\begin{equation}
    \partial_i \omega_i(r) =  \Omega(r)\partial_i n_i(r) + n_i \partial_i \Omega(r) =0
\end{equation}

The  Euler terms will depend upon distribution of vorticity strength over the surface.

It will then be a problem to find vorticity strength by balancing Euler  terms leading to conservation of the (PDF of) circulation.

The Euler term will vanish if the $ v \omega$ term reduced to a gradient
\begin{equation}
    \omega_k(r)\left(\partial_i \delta_{k c} - \delta_{i c} \partial_k\right)   \int_{S_C} d\sigma(r') \frac{ \omega_c(r')}{4 \pi |\vec r'- \vec r|} =\partial_i H(r)
\end{equation}
where  $H(r)$ is some function on the surface. Therefore $\partial_i H(r)$ lies in the local tangent plane where the loop belongs.
So, we have an equation for vorticity strength
\begin{equation}\label{VorticityEq}
    e_{p q i} n_p(r) \partial_q \left[\omega_k(r)\left(\partial_i \delta_{k c} - \delta_{i c} \partial_k\right)\int_{S_C} d\sigma(r') \frac{\omega_c(r')}{4 \pi |\vec r'- \vec r|} \right] =0
\end{equation}

Initially, we derive that only for the points $r\subset C$, but it must also be valid everywhere inside the minimal surface. 
Here is the reason why: every point of vorticity sheet is moved by the common velocity field, given by Biot-Savart integral, so for the sheet to be stationary the same equation should be valid at every point of the surface. 

We are, in fact, applying Navier-Stokes equation for vorticity at each point of this sheet and looking for stationary  vorticity distribution. The loop equation is just a technical tool to achieve that.
Imagine a small loop $\delta C $ drawn on the surface around the point $r_1\subset S_C$ and compute the time derivative of circulation around that small loop. Repeating the same arguments as before we get the same equation on every point on the surface.

So we have to find the way to solve this integro-differential equation numerically or analytically.

This is the general plan. Now let us go to the details which we did not elaborate here.

\section{Area Derivative}

The notion of Stokes-type functionals and the corresponding area derivative is central to our theory.
This notion\footnote{We are only considering so called Abelian vector fields, where the notion of Stokes type functional simplifies.} is abstracted from the properties of velocity circulation
\begin{equation}
    \Gamma[C] = \oint_C v_i(r) d r_i
\end{equation}

If we assume that $C$ consists of some number of closed loop components and add one more infinitesimal closed component $\delta C$ to $C$ around  arbitrary point $r$ in space ten clearly the circulation is additive
\begin{equation}
    \Gamma[C+ \delta C] = \Gamma[C] + \oint_{\delta C} v_i(r) d r_i
\end{equation}
at infinitesimal loop $\delta C$
\begin{equation}
    \oint_{\delta C} v_i(r) d r_i \rightarrow \delta \sigma_{i} \omega_i(r();\, \delta \sigma_i= \frac{1}{2} e_{i j k}\oint_{\delta C} r_k d r_k 
\end{equation}

So, by definition, the functional $W[C]$ with the property
\begin{equation}
    W[C+ \delta C] \rightarrow W[C] + \delta \sigma_{i} W_i(r,C)
\end{equation}

is called a Stokes-type functional and $W_i(r,C)$ is called its area derivative $\frac{\partial W}{\partial \sigma_i(r)}$.
In addition to existence, it has to satisfy the conservation law (Stokes condition):
\begin{equation}
    \partial_i \frac{\partial W}{\partial \sigma_i(r)} =0
\end{equation}
reflecting the fact that $\partial_i \omega_i =0$.

Clearly, our PDF $P(C,\Gamma)$ being a function of circulation, belongs to the class of Stokes-type functionals with
\begin{equation}
    \frac{\partial \Gamma[C]}{\partial \sigma_i(r)} = \omega_i(r)
\end{equation}
so we must look for a stationary solution for this PDF in the Stokes-type functionals as well.

An obvious basis for such functionals would be
\begin{equation}
    W[C] = \sum_n \frac{1}{n!}\oint_C d r_{1,i_1} \dots \oint_C d r_{n,i_n}F(1,...n)
\end{equation}
where $F(1,...n)$ depends both on coordinates $r_1 \dots r_n$ and their vector indexes $i_1 \dots i_n$.
The area derivative would amount to eliminating one integration and introducing the corresponding curl
\begin{equation}
    \frac{\delta W[C]}{\delta \sigma_i(r)} = e_{i j k} \partial_j \sum_n \frac{1}{n!}\oint_C d r_1 \dots \oint_C d r_n F(1,...n,n+1);\, r_{n+1} = r, i_{n+1} = k
\end{equation}

However, this is not the only way to represent the Stokes-type functional. The minimal area $A_C$ is another example. It has to be as the area law  for the Wilson loop $W[C] = \left<\hat{T}\exp(\oint A_\mu d x_\mu)\right> \rightarrow \exp(-A_C)$ is known to represent an asymptotic solution at large loops in Non-Abelian gauge theories (and in Abelian in lower dimensions as well).

In fact, the minimal area is a limit of a Stokes-type functional,  which we call Regularized area
\begin{equation}
    A^{Reg}_C =\min_{S_C} \int_{S_C} d \sigma_{i}(r_1) \int_{S_C} d \sigma_{j}(r_2) \delta_{i j} \Delta(r_1-r_2)
\end{equation}
with 
\begin{align}
&\Delta(r) = \frac{1}{r_0^2} \exp\left(-\pi \frac{r^2}{r_0^2}\right) \\
&r_0 = \left(\frac{\nu^3}{\mathcal{E}}\right)^{\frac{1}{4}}
\end{align}
The area derivative of $A^{Reg}_C$ can be computed by adding one more little surface $\delta S$ around $\delta C$ and noting that now the area is not additive : there is a cross term
\begin{equation}
    A^{Reg}_{C + \delta C} = \int_{S_C + \delta S} d \sigma_{i}(r_1) \int_{S_C + \delta S} d \sigma_{j}(r_2) \delta_{i j} \Delta(r_1-r_2)  \rightarrow
    A^{Reg}_{C} + 2 \delta \sigma_{i}(r) \int_{S_C} d \sigma_{j}(r_2) \delta_{i j} \Delta(r-r_2) 
\end{equation}
so that
\begin{equation}
    \frac{\delta A^{Reg}_C}{\delta \sigma_i(r)} = 2  \int_{S_C} d \sigma_{j}(r_2) \delta_{i j} \Delta(r-r_2)
\end{equation}
In virtue of minimality the variation of $A^{Reg}_C $ when the surface $S_C$ changes to $S'_C $ will be the surface integral over little closed surface $\delta S$ between  $S'_C$ and $S_C $ 
\begin{equation}
   \delta A^{Reg}_C = \oint_{\delta S} d \sigma_i(r) \frac{\delta A^{Reg}_C}{\delta \sigma_i(r)}\rightarrow \delta V \partial_i \frac{\delta A^{Reg}_C}{\delta \sigma_i(r)} =0
\end{equation}
Here $\delta V$ is the volume inside $\delta S$. This variation must vanish in virtue of minimality of regularized Area.

One can directly verify the Stokes condition for the minimal surface.
Let us choose $x,y$ coordinates in a local tangent plane to the minimal surface at some point which we set as an origin.  In the quadratic approximation (which will be enough for our purpose), equation of the surface reads:
\begin{align}
     z &= \frac{1}{2} \left(K_1 x^2  + K_2 y^2\right)\\
     n &= \frac{\left[ -K_1 x, -K_2 y, 1\right]}{\sqrt{1 +  K_1^2 x^2 + K_2^2 y^2}}\label{Normal2}
\end{align}
where $K_1, K_2$ are main curvatures in planes $y=0, x=0$.

Now, at $r_0 \rightarrow 0$ the Stokes condition reduces to :
\begin{align}
    &\partial_i \frac{\delta A^{Reg}_C}{\delta \sigma_i(0)} \propto \int_{S_C} d \sigma_i(r) \partial_i \Delta (\vec r)\\
    &=\int_{S_C} d \sigma(r) n_i(r)\partial_i \Delta (\vec r)\\
     &\propto\int_{-\infty}^{\infty}d x \int_{-\infty}^{\infty} d y \partial_z \Delta (\vec r)\\
     &\propto
     \int_{-\infty}^{\infty}d x \int_{-\infty}^{\infty} d y \frac{z}{r_0^4} \exp\left(-\pi \frac{x^2+ y^2}{r_0^2}\right)\\ &\propto
     \int_{-\infty}^{\infty}d x \int_{-\infty}^{\infty} d y \frac{K_1 x^2 + K_2 y^2}{r_0^4} \exp\left(-\pi \frac{x^2+ y^2}{r_0^2}\right) \\
     &\propto K_1 + K_2 =0\label{MeanCurvature}
\end{align}
 
This is the mean curvature. So, the Stokes condition is \textbf{equivalent} to the equation for the minimal surface.

Now we can use Biot-Savart law to recover velocity field corresponding to vorticity defined  as area derivative of regularized minimal area.
\begin{equation}
    V_i(r) = e_{i j k}\int d^3 r' \frac{r'_j - r_j}{4 \pi |r-r'|^3} \partial_i \frac{\delta A^{Reg}_C}{\delta \sigma_i(r')} ;\, \partial_i V_i =0, e_{i j k} \partial_i V_j = \partial_i \frac{\delta A^{Reg}_C}{\delta \sigma_i(r)}
\end{equation}

now we can verify that in virtue of the Stokes theorem the circulation for this velocity reduces to the area:
\begin{align}
    &\oint_C d r_i V_i(r) = \int_{S_C} d \sigma_i(r)  \frac{\delta A^{Reg}_C}{\delta \sigma_i(r)}  \\
    &= \int_{S_C} d \sigma_i(r)  2 n_i(r) = 2\int_{S_C} d \sigma(r) = 2 A_C
\end{align}

\section{Viscosity and Boundary condition}

There is an important feature of the viscosity term in the loop equation to be understood before we proceed any further. Namely, it involves the derivatives of the vorticity along the surface in the normal direction to external loop. This translates into the corresponding derivative of the $\Omega(r)$,  which is in fact, singular. 
The viscosity term has the structure
\begin{equation}\label{ViscosityTerm}
    \nu e_{j i k} \oint_C d r_i  \partial_j \omega_k 
\end{equation}
which involves normal vector $\hat{y}$ to the loop along the surface
\begin{equation}
    \hat{y} \propto  d r \times n
\end{equation}
With our Area Anzatz $\omega_k = n_k \Omega$ this would lead to derivative of $\Omega$ inside the surface at its boundary. But this is where $\Omega$ has a step function singularity switching from finite value inside to zero outside. One may think that the inside limit of derivative should be taken, but careful analysis with regularized Area show that this is wrong.
\begin{align}
      &\partial_y  \frac{\delta A^{Reg}_C}{\delta \sigma_z(0)} \\
      &\propto \int_{S_C} d \sigma(r) \Omega(r)\partial_y \Delta \left(\vec r\right)\\
      &\propto  \int_{-\infty}^{\infty} d x \int_0^\infty d y \Omega(x,y) \partial_y \Delta\left({x,y}\right)\\
      &\propto  \int_{-\infty}^{\infty}d x \int_{0}^{\infty} d y \frac{\Omega(0)y + \partial_y \Omega(0) y^2}{r_0^4} \exp\left(-\pi \frac{x^2+ y^2}{r_0^2}\right)\\ 
     &\propto \frac{\Omega(0)}{r_0}  + \partial_y \Omega(0)
\end{align}

As a result, the leading term diverges and reduces to perimeter with extra factor $1/r_0$  rather than the area:
\begin{equation}
    \frac{\nu}{r_0} \oint d \theta \left|\vec C'(\theta)\right|\Omega\left(C(\theta)\right)
\end{equation}

This term would not diverge if $\Omega$ vanishes at the boundary, or in general case, is a total derivative of periodic function. 
\begin{equation}
    \Omega\left(C(\theta)\right) =  \frac{\Phi'(\theta)}{\left|\vec C'(\theta)\right|}
\end{equation}
In that case we would be left with finite term
\begin{equation}
 \nu e_{j i k} \oint_C d r_i n_k \partial_j \Omega(r)
\end{equation}

\section{Euler Loop Equation}

Let us start with the Euler loop equation, we add viscosity later.

The fixed point (i.e. stationary solution) of the loop equation was shown to be a  following function:
\begin{equation}\label{Scaling}
    P(C,\Gamma) = A_C[\Omega]^{-\frac{1}{2}}\Pi\left(\Gamma  A_C[\Omega]^{-\frac{1}{2}}\right)
\end{equation}
where  $A_C[\Omega]$ is net vorticity over the surface
\begin{equation}\label{Area}
    A_C[\Omega] = \int_{S_C} d \sigma(r) \Omega(r)
\end{equation}
where local vorticity strength $\Omega(r)$ is some external field defined on the minimal surface. 
This surface satisfies usual equation 
\begin{equation}\label{meanCurv}
    K_1(r) + K_2(r) =0
\end{equation}
where $K_1(r), K_2(r)$ are two principal plane curvatures at the surface at the point $r$.
The area derivative yields
\begin{equation}
    \frac{\delta A_C[\Omega]}{\delta \sigma_i(r)} = 2 n_i(r) \Omega(r)
\end{equation}

We start with stationary loop equation in the WKB approximation 
\begin{equation}
     \Pi(\gamma)\propto exp\left(-S(\gamma)\right); S\rightarrow \infty
\end{equation}
\begin{equation}
    r_0\left(\frac{\partial S}{\partial {A_C}}\right)^2\oint_C d r_i  \int_{S_C} d\sigma(r')\frac{\delta A_C}{\delta \sigma_k(r)} \frac{\delta A_C}{\delta \sigma_c(r')}\left(\partial_i \delta_{k c} - \delta_{i c} \partial_k\right)\frac{1}{4 \pi |\vec r'- \vec r|} =0
\end{equation}
By rewriting the equation
\begin{equation}
    \oint_C d r_i F_i(r) =0
\end{equation}
for a loop $C$ on a surface as condition of vanishing surface curl 
\begin{equation}
    e_{p q i} n_p(r) \partial_q F_i(r) =0
\end{equation}
we arrive to an integral equation for $\Omega$
\begin{equation}\label{OmegaEq}
    e_{p q i} n_p(r) \partial_q \left[\Omega(r) n_k(r)\left(\partial_i \delta_{k c} - \delta_{i c} \partial_k\right)  \int_{S_C} d\sigma(r') \frac{\Omega(r') n_c(r')}{4 \pi |\vec r'- \vec r|}\right] =0
\end{equation}

This equation is exact -- no approximations were made except neglecting viscosity as well as random external forces and taking the limit of infinitely thin viscosity sheet. 
In other words, this is the equation for the stationary PDF of circulation in inertial range $\Gamma \gg \nu, |C| \gg r_0$, with $\mathcal {E}$ being Kolmogorov's energy dissipation rate. 

As it was noted in the previous paper, with flat loop when the minimal surface is flat as well, the normal vector is constant along the surface. In this case the constant $\Omega$ solves this equation, as it reduces to
\begin{equation}
      e_{p q i} n_p n_k n_c\partial_q\left(\partial_i \delta_{k c} - \delta_{i c} \partial_k\right) \int_{S_C} d\sigma(r') \frac{1}{4 \pi |\vec r'- \vec r|} =0
\end{equation}
However, the constant $\Omega$ does not satisfy the boundary condition of vanishing $\int d l\Omega$. Apparently, this solution is not valid near the boundary $r\rightarrow C$, where the integrals are singular (but still finite).

So, correct solution may be constant inside the surface and reduce to $\partial_l \Phi $ at thin layer near the boundary.

\section{Small Curvature Regime}

As we already know, for a flat surface any constant $\Omega(r)$ represents a solution. Let study the case of small curvature of the surface and
rewrite exact equation to single out the curvature terms

\begin{align}
    &e_{p q i} n_p(r) K_{q k l}(r) n_l(r) \Omega(r) \left(\partial_i \delta_{k c} - \delta_{i c} \partial_k\right)  \int_{S_C} d\sigma(r') \frac{\Omega(r') n_c(r')}{4 \pi |\vec r'- \vec r|} +\\
    &e_{p q i} n_p(r)   \left[\partial_q\Omega(r)\right] \left(n_c(r) \partial_i  - \delta_{i c} n_k(r)\partial_k\right)  \int_{S_C} d\sigma(r') \frac{\Omega(r') n_c(r')}{4 \pi |\vec r'- \vec r|} +\\
    &e_{p q i} n_p(r) n_k(r)\Omega(r) \partial_q \left(\cancel{\partial_i \delta_{k c}} - \delta_{i c} \partial_k\right)  \int_{S_C} d\sigma(r') \frac{\Omega(r') n_c(r')}{4 \pi |\vec r'- \vec r|} =0
\end{align}
The  terms remaining after all cancellations
\begin{align}
&e_{p q i} n_p(r) K_{q k l}(r) n_l(r) \Omega(r) \left(\partial_i \delta_{k c} - \delta_{i c} \partial_k\right)  \int_{S_C} d\sigma(r') \frac{\Omega(r') n_c(r')}{4 \pi |\vec r'- \vec r|} +\\
    &e_{p q i} n_p(r)   \left[\partial_q\Omega(r)\right] \left(n_c(r) \partial_i  - \delta_{i c} n_k(r)\partial_k\right)  \int_{S_C} d\sigma(r') \frac{\Omega(r') n_c(r')}{4 \pi |\vec r'- \vec r|} -\\
    &e_{p q i} n_p(r) n_k(r)\Omega(r) \partial_q  \partial_k  \int_{S_C} d\sigma(r') \frac{\Omega(r') n_i(r')}{4 \pi |\vec r'- \vec r|} =0
\end{align}
are  suitable for iterations starting from $\Omega(r) = 1$. It has the form
\begin{align}
& A_{q}(r) \partial_q\Omega(r)  = \Omega(r)B(r);\\\label{AOmegaB}
& A_q(r) = e_{p q i} n_p(r) \left(n_c(r) \partial_i  - \delta_{i c} n_k(r)\partial_k\right)  \int_{S_C} d\sigma(r') \frac{\Omega(r') n_c(r')}{4 \pi |\vec r'- \vec r|};\\
& B(r) = e_{p q i} n_p(r)  \left(K_{q k l}(r) n_l(r) + n_k \partial_q\right) \left(-\partial_i \delta_{k c} + \delta_{i c} \partial_k\right)  \int_{S_C} d\sigma(r') \frac{\Omega(r') n_c(r')}{4 \pi |\vec r'- \vec r|}
\end{align}
In the flat limit when $n_i(r) \rightarrow Z_i = (0,0,1)$
\begin{align}\label{A0}
    A_\alpha(r) \rightarrow e_{\alpha \beta}   \partial_\beta   \int_{S_C} d\sigma(r') \frac{\Omega(r')}{4 \pi |\vec r'- \vec r|}
\end{align}
As for the $B$ term, it vanishes in flat limit, as we have already seen but we need just the leading term in $B$, proportional to gradients of normal vector (which is the curvature quadratic form).
So, we are left with local equation
\begin{equation}\label{AdOmegaB}
     A_q(r) \partial_q \ln \Omega(r)  = B(r) ;
\end{equation}
This equation can be solved in general form, as we know from the college math course. Let us introduce the path $P_i(\alpha)$ on our surface  along the direction of vector $A_i(r)$
\begin{equation}
    P'_i(\alpha) = A_i\left(P(\alpha)\right);\, P(0) \in C; 
\end{equation}
Then, the equation (\ref{AdOmegaB}) becomes an ODE:
\begin{equation}
    \frac{d \ln \Omega\left(P(\alpha)\right)}{d \alpha} = B\left(P(\alpha)\right)
\end{equation}
which can be solved:
\begin{equation}\label{OmegaPath}
    \Omega\left(r\right) = \Omega(C) \exp\left(\int d \alpha \, B\left(P(\alpha)\right)\right)
\end{equation}

Let us be more specific about boundary conditions. Both $A_q(r), B(r)$ have weak singularities when $ r \rightarrow C$, so that $\int A_i d \alpha $ and $ \int B d \alpha$ converge. So, the path $P_i(\alpha)$ can originate at arbitrary point $C(\theta_0)$ and go from there inside the surface \footnote{exactly at the boundary the vector $A_i$ is directed along the loop but it turns inside the surface after that, as the logarithic singularity in tangent component of $A_i$ is integrable.}. This leaves the boundary value $\Omega(C(\theta_0))$ arbitrary. For each point at the boundary there will presumably be a path leading inside the surface. If this path $P_{r_1 r_2}$ starts at $r_1 \in C$ ends on another point $r_2 \in C$, we have an ambiguity: which of two paths $P_{r_1 r}, P_{r_2 r}$ leading to $r$  to use to compute $\Omega(r)$. This ambiguity would disappear if 
\begin{equation}\label{Omega12}
     \Omega\left(r_2\in C\right) = \Omega\left(r_1\in C\right) \exp\left(\int d \alpha \, B\left(P_{r_1 r_2}(\alpha)\right)\right)
\end{equation}
for every such pair of endpoints of our paths.

Using the  viscosity boundary condition we should have
\begin{equation}\label{OmegaIntegral}
    \Omega(r) = \frac{\Phi'(\theta)}{\left|\vec C'(\theta)\right|}\exp\left(\int d \alpha \, B\left(P_{C(\theta) r}(\alpha)\right)\right)
\end{equation}

At this point we do not have any explicit expression for the boundary value $\Phi'(\theta) $. The relation (\ref{Omega12}) connects pairs of values of $\Phi'(\theta)  $ at the two ends of each path. Once we know the solution inside the surface, we can try to use this linear set of equations to remove the ambiguity in $\Phi $. 

Let us note that both $A_i(r), B(r)$ are odd with respect to reflection of the normal vector ( corresponding to reflection of orientation of the surface). However our vorticity density $\Omega(r)$ remains invariant, as both terms of equation change sign.
Note also that our equations are scale invariant: they can determine $\Omega(r)$ up to the overall scale factor.  Rescaling of $\Omega$ leads to change of the scale factor in relation $\Gamma^2 \propto A_C$. In terms of conventional dimensional counting $\Omega \propto 1/T$ as it measures mean vorticity on the surface. The time scale drops from the stationary Euler dynamics which is why the scale factor in $\Omega$ remains undetermined.

The expansion around flat surface corresponds to iterations of this equation starting with $\Omega = 1$. The first correction would use $A_i(r)$ and $B(r)$ with $\Omega(r')=1$ inside the integrals and all integration going over the flat surface with leading curvature correction. This first approximation already satisfies the viscosity boundary condition.

\section{Triangulated Minimal Surface and Numerical Approach}

\begin{figure}[p]
    \centering
    \includegraphics{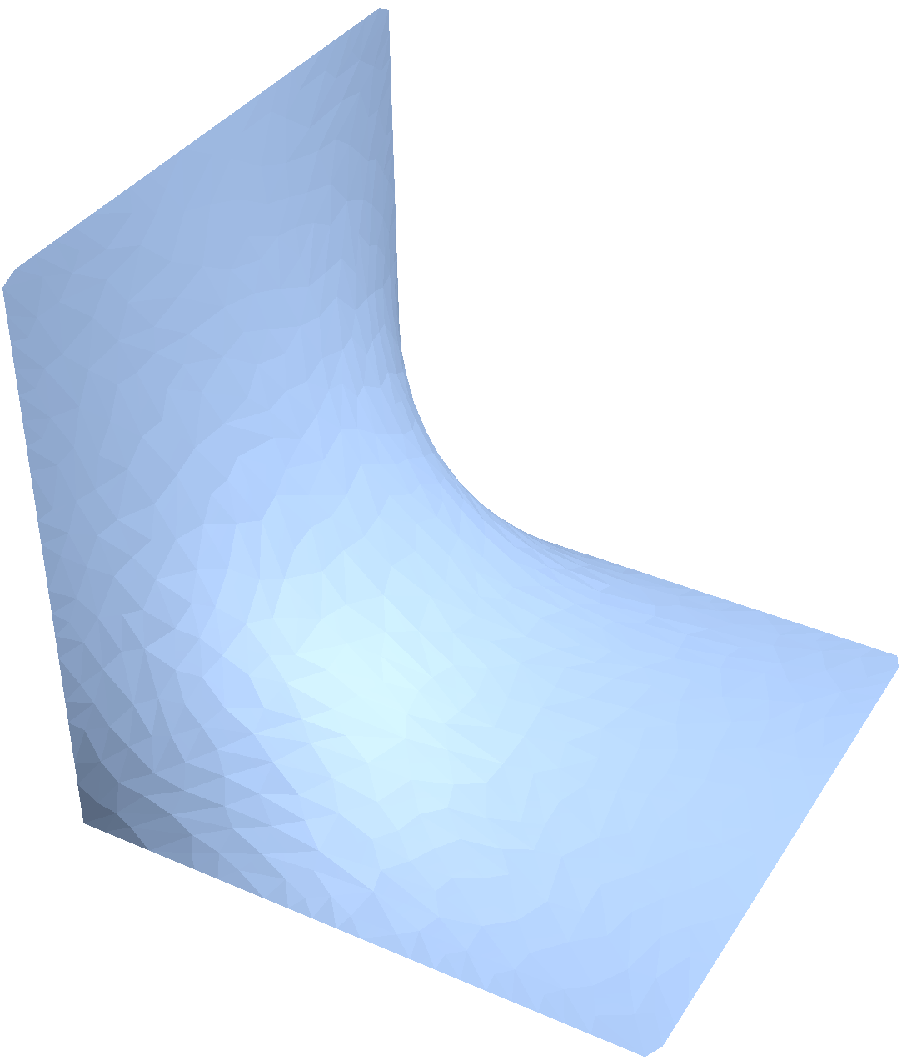}
    \caption{Minimal surface bounded by "soccer gates" approximated by triangulation.}
    \label{fig:SoccerGates}
\end{figure}

We propose the following numerical solution of the Euler loop equation. 
For any given closed loop $C$ we can use the \Mathematica code from \cite{MinSurfaceX} to approximate the minimal surface as a triangulated surface $\mathcal{T}_C$ bounded by a polygon approximating $C$. (see Fig. \ref{fig:SoccerGates}).

The normal vectors will be associated with the centers $r_i $ of these triangles $T_i$ and and so will be the $\Omega$ field. We shall have an array $\Omega(r_i)$ one for each triangle $r_i \in \mathcal{T}_C$. The surface integrals will become sums over triangles with $\Delta \sigma(r)$ being represented by area of each triangle. The line integral $\int_C | d\vec r| \Omega(r)$ will become the sum over edges of the approximating polygon, with $\Omega(r)$ corresponding to the boundary triangles with these edges.

It is understood that these edges as well as the edges of inner triangles in triangulated minimal surface, would have the viscous scale $r_0$, and the number $N$ of triangles will play the role of the Reynolds number, going to infinity. The Biot-Savart kernel will only enter at the centers of triangles, which naturally cuts off the divergencies at $r'\rightarrow r$ in the surface integrals. These integrals will be replaced by sums over triangles $r'$ with exception of triangle $r$. This corresponds to the principal value prescription for the continuous theory.

Now that we associate every field $n_k(r), \Omega(r)$ as well as Biot-Savart kernel $ K_i(r,r')$ with the centers of triangles we can define discrete approximation of the Loop Equation as the following relation. At each  triangle $r$ with directed clockwise edges $E_1,E_2,E_3$ , there could be one, two or three adjacent triangles, sharing an edge with $r$.
So, for each edge $E$ we define
\begin{equation}\label{Sym}
    F(E) = \frac{F(r) + F(r')}{2}
\end{equation}
if this edge is shared by two triangles $r, r'$ and
\begin{equation}
    F(E) = \frac{F(r)}{2}
\end{equation}
in case this edge is a boundary edge belonging  to polygon approximating $C$.

The loop equation for circulation around the triangle $r$
\begin{equation}
    \oint_{C(r)} d \vec r \vec F(r) =0
\end{equation}
becomes in our discrete version
\begin{equation}\label{discreetLoop}
    \sum_{t=0}^3 \vec E_t \vec F(E_t) =0
\end{equation}

Here 
\begin{equation}
    F_i(r) = \Omega(r) \sum_{r'\ne r} \Delta \sigma(r') K_i(r,r')\Omega(r')
\end{equation}
\begin{align}\label{Kirr}
    &K_i(r,r') =  \frac{ n_i(r') \rho_k n_k(r) - n_k(r)n_k(r') \rho_i }{4 \pi |\vec \rho|^3}\\
    & \rho_i = r_i - r'_i
\end{align}
with the sum $\sum_{r'\ne r} $ going over all triangles $ r'$ of the surface $\mathcal{T}_C $ with areas $ \Delta \sigma(r')$ except the one with center $r$. 

The symmetrization (\ref{Sym}) is not just a computational trick to eliminate first order terms in local limit. With this symmetrization the discrete version of circulation over triangle is additive in the sense of the Stokes theorem: adding together two such circulations with adjacent path traversed in opposite directions , like in two adjacent triangles, the two contributions of this adjacent path cancel each other, so that we get the circulation of the joined loop (rhombus in case of two adjacent triangles). Repeating this for all adjacent triangles in $\mathcal{T}$ we get the total circulation over the polygon approximating $C$, in the same way as we would get it via Stokes theorem in continuous theory.

In other words, with this symmetrization we have discrete analog of the Stokes theorem, making the sum of circulations over all triangles in $\mathcal{T}$ to reduce to circulation over the bounding polygonal loop. This is important, as the errors in the Stokes formula inside the surface would accumulate and produce the area term, much larger then correct answer which is circulation over the bounding loop, scaling as perimeter.

Note however, that the internal term $F(r)$ in (\ref{Sym}) drops out after summation of all three terms in (\ref{discreetLoop}), as the vector sides $\vec E_t $ of triangle add to to zero vector. So, this symmetrization is needed just to avoid accumulation of numerical errors for large surface.

There will be the same number of equations as there are independent variables $\Omega(r)$.
We should solve these equations along with boundary condition 
\begin{equation}\label{boundaryCond}
    \sum_{E \in C} \Omega(r) |\vec E| =0
\end{equation}
where the sum goes over edges in  $C$ and $r$ is the boundary triangle with this edge.
The number of equations therefore is $N+1$ for $N$ triangles on a surface. Extra equation means that we have to minimize the sum of squares of all the $N$ equations for $N-1$ independent variable left after solving the boundary condition.

In fact, there are  $N-2$ independent variables as the common scale of $\Omega $ array is not determined by these scale invariant equations.
The normalization of $\Omega $ is provided by its relation to $A_C$
\begin{equation}\label{normalizationOmega}
    A_C = \sum_{r \in \mathcal{T}} \Delta \sigma(r) \Omega(r)
\end{equation}

At finite number of triangles this set of equations can be solved by \Mathematica, with gradients and Hessian computed analytically. We did it here for small enough number of $1194$ triangles to illustrate the method. For the purposes of comparison with turbulence simulation the number of triangles should be set to  $\sim 10^6$ where the surface becomes smooth at the level of viscosity scale. The same \Mathematica program \cite{MLE2019}  would do it, with compiled $C$ code exported to supercomputer.

The target function minimized was sum of squares of all triangle equations (\ref{discreetLoop}) plus smoothness $\left(\nabla \Omega\right)^2 $ term
 \begin{equation}
    \sum_{r \in \mathcal{T}} \left[\left(\sum_{t=0}^3 \vec E_t \vec F(E_t)\right)^2 + \Delta \sigma(r) \sum_{k=1}^3 \frac{\left(\Omega(r) - \Omega(r_k)\right)^2 \Delta \sigma(r_k)}{\left|\vec r - \vec r_k\right|^2}\right]
 \end{equation}
where $r_k$ are three neighbors of triangle $r$ with centers at $\vec r_k, \vec r$. 

\begin{figure}[p]
    \centering
    \includegraphics{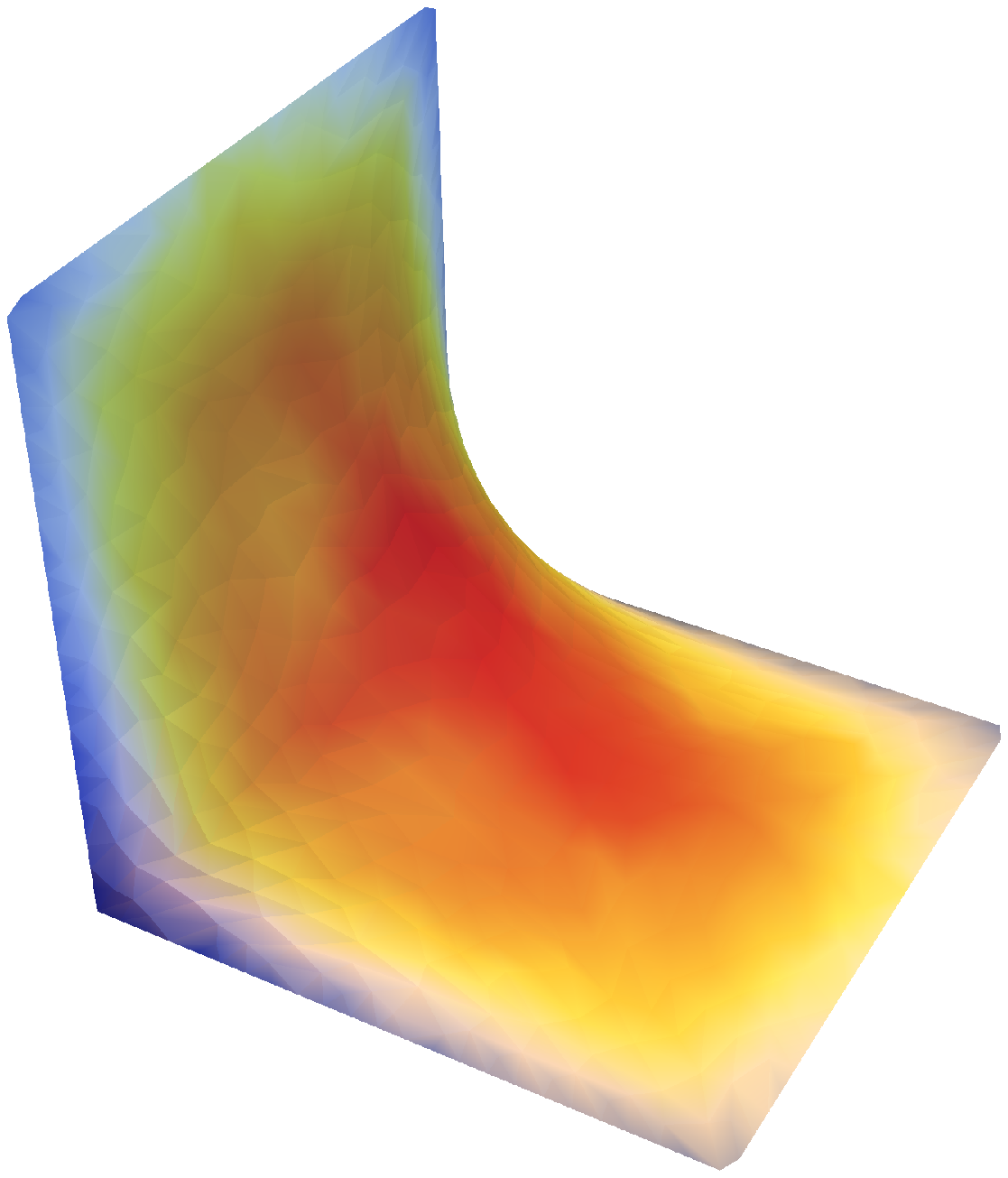}
    \caption{The Omega Field on Minimal Surface, shown by the color in Temperature Map from blue to red.}
    \label{fig:OmegaSurface}
\end{figure}

The boundary condition (\ref{boundaryCond}) and normalization $A_C =1$ was added as constraint. The resulting $\Omega$ field varies from negative values at upper corners to extreme positive values in the center, with values at the middle corners being extreme negative and those at the lower corners being positive. The $\Omega$ field was attributed to vertices by averaging over surrounding triangles, and then displayed as interpolated color inside each triangle.
(see Fig. \ref{fig:OmegaSurface}).

The $\Omega$ field is scale invariant: with rescaling of the coordinates of the surface by the same factor $ L$ the area scales as $L^2$ but the $\Omega $ field would not change if we normalize it to $A_C = A$, where $A$ is geometric area of the surface. This follows from the fact that the kernel (\ref{Kirr}) scales as $1/L^2$ compensating the scale of $\Delta \sigma(r) \sim L^2$, making $F_i(r)$ as well as our target scale invariant, as well as the boundary condition and normalization $A_C = A$.

\section{Viscosity Correction}

Let us now study the linear correction coming from viscosity term in the loop equation, again in the leading WKB approximation with small PDF $ P(C,\Gamma) = \exp\left(-S(A_C,\Gamma)\right)$.
We look for the following viscosity correction for PDF
\begin{equation}
    P(C,\Gamma) = \exp\left(-S(A_C,\Gamma)\right)\left(1 + \frac{\nu}{r_0} V(A_C,\Gamma) \int d \sigma(r) \Omega_1(r)\right)
\end{equation}
Here $V(A_C,\Gamma)$ and $\Omega_1(r)$ are the functions to be found from the perturbed loop equation.

Let us collect corrections to the Euler Loop equation (\ref{OmegaEq}).
Taking out the common factor $\exp\left(-S(A_C,\Gamma)\right)$ we have the following terms inside the loop integral $\oint_C d r_i$
\begin{align}
    &2\nu \partial_\Gamma S \partial_{A_C} S e_{j i k}  n_k(r) \partial_j \Omega(r)\\
    &-4\nu \partial_{A_C} S V(A_C,\Gamma)  \int_{S_C} d\sigma(r')\left(\Omega(r)\Omega_1(r') + \Omega_1(r)\Omega(r')\right) K_i(r,r')
\end{align}
with $K_i(r,r')$ defined in (\ref{Kirr}). \footnote{There are also terms with variations of $V(A_C,\Gamma)$, but these terms cancel among themselves in virtue of the Euler loop equations.}
This must be zero modulo gradient $\partial_i H(r)$ to produce zero after integration over the loop.
Let us compare the dependence of $\Gamma, C $ first.
Taking out common factor $\partial_{A_C} S$ we observe that the terms will match provided
\begin{equation}
    \partial_\Gamma S  = V(A_C,\Gamma)
\end{equation}
After that, we get linear integral equation for the $\Omega_1(r)$
\begin{equation}
    e_{p q i} n_p(r) \partial_q \left[-e_{j i k}  n_k(r) \partial_j \Omega(r) + 2\int_{S_C} d\sigma(r') \left(\Omega(r)\Omega_1(r') + \Omega_1(r)\Omega(r')\right) K_i(r,r')\right] =0
\end{equation}

We leave for future numerical studies of this equation, which can be discretized on a triangulated minimal surface.
As for the moments of PDF , the correction goes as follows
\begin{align}
    &\Delta \left< \Gamma^p \right> \propto  \\
    &\int_{-\infty}^{\infty} d \Gamma \Gamma^p\exp\left(-S(\Gamma,C)\right) \partial_\Gamma S(\Gamma,C) =\\
    &\int_{-\infty}^{\infty} d \Gamma p\Gamma^{p-1}\exp\left(-S(\Gamma,C)\right) = p \left< \Gamma^{p-1} \right> 
\end{align}

\section{Conclusions}

What can we conclude from this analysis? 
\begin{itemize}
\item The deviation from the flat loop changes the area law, but this change is calculable analytically for small deviations. The minimal area is replaced by the integral over the minimal surface of some vorticity density $\Omega(r)$, which satisfies an integral equation (\ref{OmegaIntegral}) on a minimal surface.

\item The whole problem of computing $\Omega(r)$ can be discretized on a triangulated surface, which can be built in \Mathematica to approximate the minimal surface with high accuracy for arbitrary loop. The discretized set of equations for the values of $\Omega(r)$ in the centers of triangles can also be solved by means of \Mathematica minimization program \cite{MLE2019} at any given triangulated minimal surface.

\item The most striking prediction of this theory is that the mean vector vorticity at the loop $C$ is proportional to $\Omega(r) \vec n(r)$, where both factors are calculable for any given $C$, in particular for the Soccer Gate loop.

\item The leading viscosity correction term mixes the moments $\left< \Gamma^p \right>$  with $\left< \Gamma^{p-1} \right>$ resembling the bi-fractal behavior observed in \cite{S19} and explicitly breaking the time reversal symmetry.
\end{itemize}
 
 \section{Acknowledgements}
 I am grateful to Sachin Bharadwaj, Kartik P. Iyer, Sasha Polyakov, Katepalli R. Sreenivasan, Victor Yakhot and members of Courant Institute Seminar for stimulating discussions. 
 
\bibliographystyle{unsrt}  


\end{document}